\documentclass{Interspeech}
\interspeechcameraready 

\title{Gaze-Enhanced Multimodal Turn-Taking Prediction in Triadic Conversations}

\author[affiliation={1,2*}]{Seongsil}{Heo} 
\author[affiliation={2}]{Calvin}{Murdock}
\author[affiliation={2}]{Michael}{Proulx}
\author[affiliation={2}]{Christi}{Miller}

\affiliation{}{University of California, Santa Cruz}{USA}
\affiliation{}{Meta Reality Labs Research}{USA}
\email{sheo1@ucsc.edu, \{cmurdock,michaelproulx,christim\}@meta.com}

\begin{document}
\maketitle
\begingroup
\renewcommand\thefootnote{}\footnote{* Work done during an internship at Meta.}
\addtocounter{footnote}{-1}
\vspace{-10mm}
\endgroup

\begin{abstract}
%3rd version
Turn-taking prediction is crucial for seamless interactions. This study introduces a novel, lightweight framework for accurate turn-taking prediction in triadic conversations without relying on computationally intensive methods. Unlike prior approaches that either disregard gaze or treat it as a passive signal, our model integrates gaze with speaker localization, structuring it within a spatial constraint to transform it into a reliable predictive cue. Leveraging egocentric behavioral cues, our experiments demonstrate that incorporating gaze data from a single-user significantly improves prediction performance, while gaze data from multiple-users further enhances it by capturing richer conversational dynamics. This study presents a lightweight and privacy-conscious approach to support adaptive, directional sound control, enhancing speech intelligibility in noisy environments, particularly for hearing assistance in smart glasses.

\end{abstract}
%\renewcommand{\thefootnote}{\fnsymbol{footnote}}
%\footnote{Work done during an internship at Meta Reality Labs Research.}
%\renewcommand{\thefootnote}{\arabic{footnote}}

\section{Introduction}

Turn-taking is a fundamental mechanism in triadic conversations, enabling smooth speaker transitions and preventing interruptions. Accurate turn-taking prediction enhances conversational flow by ensuring timely responses and sustained engagement~\cite{sacks1978simplest, duncan1974structure,  riest2015anticipation}.

\begin{figure}[t]
\centering
    \resizebox{0.88\columnwidth}{!}{\includegraphics{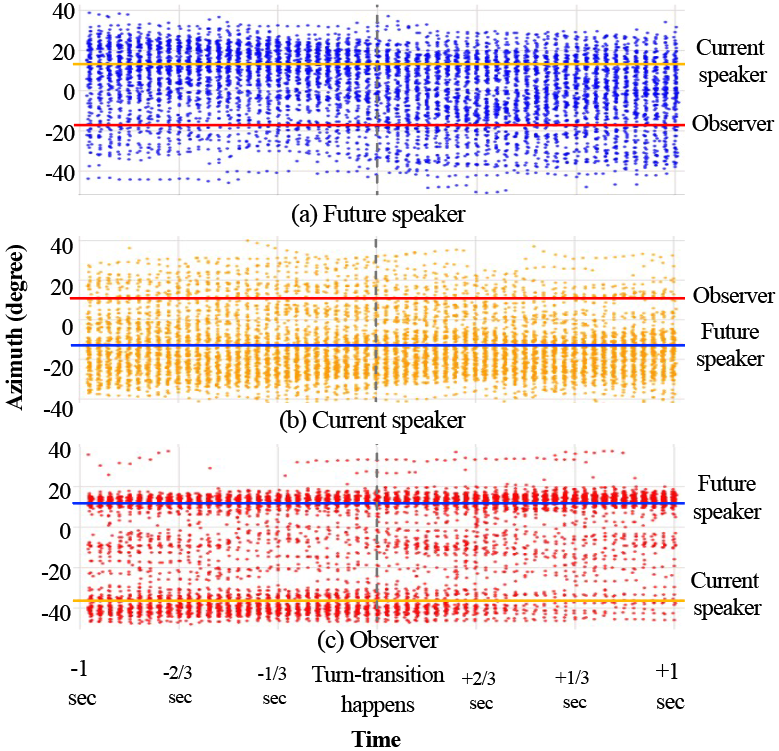}}
    \caption{Gaze behavior during turn transitions in triadic conversations is shown in peri-stimulus time histogram, with gaze azimuth coordinates around the turn-transition moment (dotted vertical lines), accumulating data around these events. This visualization is based on one session from our collected dataset, spanning 1 second before and after the transition.  Horizontal bold colored lines indicate other users' positions. For detailed descriptions of (a)-(c), please refer to the main text.} 
    \label{fig:gaze PSTH}
    \vspace{-2mm}
\end{figure}

Non-verbal cues, particularly eye gaze, play a crucial role in regulating turn transitions~\cite{degutyte2021role, jokinen2013gaze, kendrick2023turn, hirvenkari2013influence}. As shown in Fig.~\ref{fig:gaze PSTH}, (a) the future speaker often gazes at the current speaker before the transition, then either maintains or averts their gaze, becoming disengaged from the current speaker after the transition. It indicates complex response planning, or maintaining gaze for a simpler, immediate reply. Meanwhile, (b) the current speaker may direct their gaze toward the future speaker during transitions, signaling an intention to yield the floor. (c) Observers also tend to shift their gaze from the current speaker to the future speaker after the transition, anticipating the next turn.

Existing machine learning models for turn-taking prediction have primarily relied on modalities such as natural language processing (NLP) or image data~\cite{ekstedt2020turngpt, chang2022turn, kurata2023multimodal}, which require high computational resources and raise privacy concerns. In contrast, gaze-based approaches offer a lightweight, privacy-conscious alternative. Previous studies have explored gaze in turn-taking, but often relied on hand-crafted rules or statistical analyzes~\cite{degutyte2021role, ishii2016prediction}, or models trained on discretized intervals~\cite{lee2023multimodal, wang2024predicting}, limiting real-world applicability.

To address these challenges, we propose a gaze-enhanced turn-taking prediction model that integrates minimal contextual cues, including spatial heatmaps representing detected speaker locations and binary voice activity detection (VAD). These models provide a lightweight solution for continuous turn-taking prediction, enabling more dynamic and context-aware applications in triadic conversations.

Our findings demonstrate that (1) egocentric gaze and voice activity are meaningful cues, as voice activity alone lacks sufficient context, (2) multi-user gaze data further improves prediction performance by capturing richer conversational dynamics, (3) minimal complementary inputs, such as spatial heatmaps and voice activity, enable lightweight yet efficient modeling, and (4) evaluating performance from multiple perspectives, including transition-level and group-level analyzes, offers deeper insights into turn-taking dynamics. These models were evaluated on a naturalistic, triadic conversational dataset, highlighting their effectiveness in real-world scenarios.

\begin{figure*}[t]
\centering
    \includegraphics[width=0.85\textwidth]{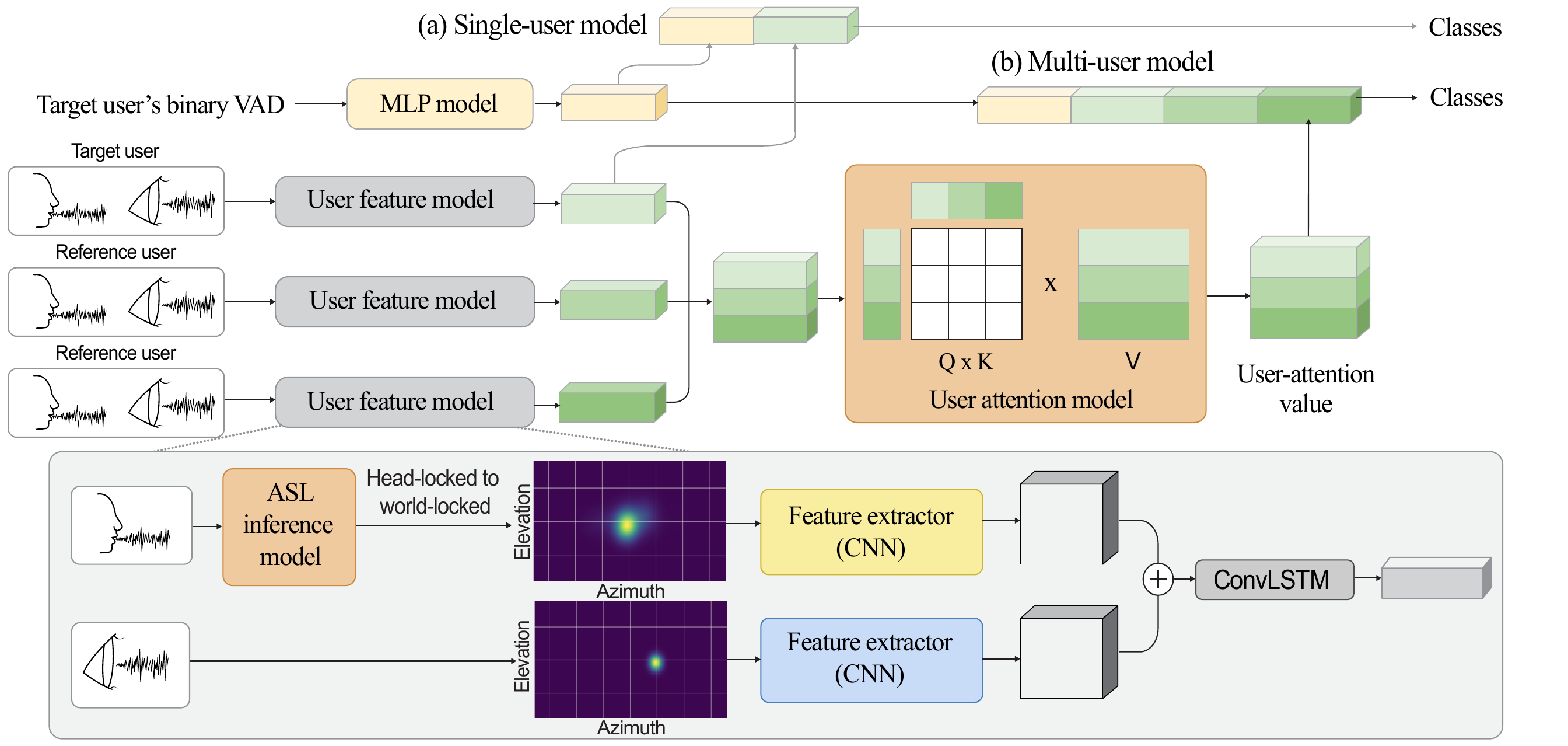}
    \caption{Architecture of the single- and multi-user models. (a) The single-user model only uses the target user's feature. (b) The multi-user model incorporates features from all users with a user attention mechanism to capture the dynamics multi-user interactions.} 
    \label{fig:model architecture}
    \vspace{-2mm}
\end{figure*}

\section{Related Work}

\subsection{Gaze-driven approaches to turn-taking prediction}
Several studies have explored the integration of gaze data for turn-taking prediction. Lee et al.~\cite{lee2023multimodal} developed a binary classification model that distinguishes between turn-taking and turn-keeping using gaze target vectors and prior conversational states. However, its dependence on pre-labeled information limits practicality in real-world applications.

Similarly, Wang et al.~\cite{wang2024predicting} proposed a model that extracts features from speech, gaze, and personality data. While this method enriches representation, it increases complexity. Moreover, their evaluation is confined to specific data segments rather than continuous sequences, which restrict its applicability.

\subsection{Audio active speaker localization}
Jiang et al.~\cite{jiang2022egocentric}  introduced an active speaker localization model that utilizes both audio and visual data, generating heatmaps to represent the spatial positions of speakers. It demonstrated robust performance across various channel configurations, including scenarios with audio-only inputs. Although it effectively identifies active speakers, it has not been applied to turn-taking prediction, to the best of our knowledge.

Integrating this model for spatial context can enhance gaze-based turn-taking prediction. By determining whether individuals are looking at the active speaker, the model can provide crucial conversational context, improving turn-taking prediction while maintaining a lightweight framework.

\section{Dataset}
We used a subset of the Reality Labs Research Conversations for Hearing Augmentation Technology (RLR-CHAT) dataset~\cite{yin2024hearing, murdock2024self, yun2025spherical}, collected using Aria glasses~\cite{engel2023project}. The dataset consists of 30 hours triadic conversations engaged in free-form conversations around a round table. Participants aged 20 to 60 years, including those without and with mild hearing loss. Each session lasted approximately 1 hour. 

Our study focused on audio, gaze, head movement, and VAD. Audio was recorded using a 7-channel microphone array (48 kHz), while gaze data was recorded with an eye-tracking camera at 30 Hz. Visual-inertial odometry (VIO) was used to gather head orientation measurements at 1 kHz, which also provided input for SLAM-based 3-DoF gaze correction. This correction removed head-motion effects to ensure that gaze direction accurately reflected participants' intended focus points. Binary VAD was derived by isolating own-voice audio through nearfield beamforming, followed by voice classification to generate a binary indicator of speaking activity.

The target user is defined as the participant wearing the glasses for whom the turn-taking prediction is performed, while reference users are all other participants. To ensure a precise analysis of turn-taking behaviors, it is essential to synchronize users' data streams with minimal error. Even slight discrepancies can alter turn-taking labels, affecting dataset reliability. To our knowledge, no existing gaze datasets in multi-party conversational settings, other than ours, achieves this level of synchronization accuracy. 

Turn-taking was labeled using a role-based and behavior-based framework~\cite{wang2024predicting}. Labels were initially generated using a binary VAD time-based algorithm, with 4 out of the 10 sessions manually re-labeled to enhance labeling accuracy. To assign labels, we analyzed Inter-Pausal Units (IPUs) extracted from binary VAD data. An IPU is defined as a continuous segment of speech separated by pauses. We applied 0.5 second smoothing to the binary VAD signal~\cite{wang2024predicting, liu2022pause, campione2002large}, merging segments where speech resumed within this threshold. 

In the role-based classification, each participant was assigned one of three roles: main speaker (if multiple speakers overlapped, the first to start speaking was assigned this role), non-main speaker (spoke during an IPU but was not the main speaker), and observer (did not speak during an IPU).

For behavior-based classification, we considered the target user's binary VAD status and the presence of a main speaker. Silence was assigned when the VAD indicates silence. Turn-taking occurred when the VAD indicates speaking, either (1) no main speaker immediately before the target user began speaking, and the previous speaker is not target user, or (2) main speaker exists immediately before the target user began speaking, and the target user stopped speaking before the main speaker finished. Turn-keeping was assigned when the VAD indicates speaking, with no prior main speaker immediately before the target user began speaking, and the previous speaker was the target user. Back-channeling occurred when The VAD indicates speaking, with a prior main speaker immediately before the target user began speaking, but the main speaker continued speaking even after the target user started speaking.

\section{Methodology}

We propose two models for turn-taking prediction: a single-user model that focuses on the target user's gaze and audio data, and a multi-user model that incorporated gaze features from all participants. An overview of these models is presented in Fig.~\ref{fig:model architecture}.

\begin{table}[t]
  \begin{center} 
  \caption{F1 scores for role-based classification. \textbf{Bold} indicates the best, while \underline{underline} represents the second-best.}
  \label{table:role-based result}
  \resizebox{0.43\textwidth}{!}{
  \begin{tabular}{p{2.2cm} | p{0.9cm} p{0.9cm} p{0.9cm} p{0.9cm}}
    \toprule
    \textbf{Model} & \textbf{All} &\textbf{Main} & \textbf{Non-main} & \textbf{Observer} \\ 
    \midrule
     Target VAD & 0.704 & 0.761 & 0.351 & \textbf{1.000} \\
     Single-user & \underline{0.746} & \underline{0.835} & \underline{0.405} & \textbf{1.000} \\
     Multi-user & \textbf{0.765} & \textbf{0.869} & \textbf{0.428} & \textbf{1.000} \\
    \bottomrule
  \end{tabular}
  }
  \end{center}
\vspace{-5mm}
\end{table}

\begin{figure}[t]
\centering
    \includegraphics[width=225pt]{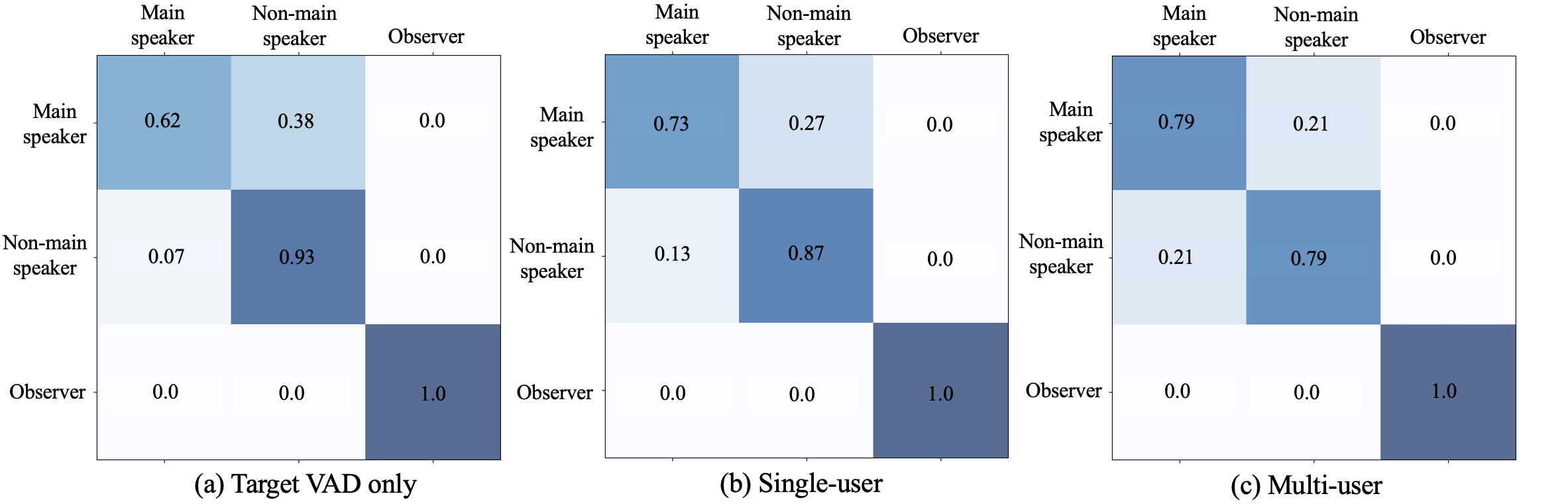}
    \caption{Confusion matrices of Table 1.} 
    \label{fig:confusion matrix}
    \vspace{-3mm}
\end{figure}

\subsection{Single-user model}
As illustrated in Fig.~\ref{fig:model architecture} (a), the single-user model processes multimodal inputs from the target user. To ensure comprehensive turn-taking analysis, each participant is assigned as the target user in turn, allowing the model to learn from multiple perspectives.

Binary VAD features are extracted using a 5-second sliding window and passed through a Multi-layer Perceptron (MLP) for feature extraction. Gaze data, represented as azimuth and elevation angles, are accumulated over a 0.2-second window (6 frames) to construct gaze heatmaps. Audio data, segmented into 0.2-second windows, is processed through an Active Speaker Localization (ASL) inference model, generating a spatial heatmap of the active speaker's location. Both heatmaps are then fed into a CNN-based feature extractor respectively.
 
Extracted gaze and audio features are combined to enhance the representation of spatiotemporal dynamics and forwarded to a Convolutional Long Short-Term Memory (ConvLSTM)~\cite{shi2015convolutional}. This produces a comprehensive user feature vector,concatenated with the binary VAD feature vector, and fed into a classifier to predict turn-taking behavior.

\subsection{Multi-user model}
The multi-user model extends the single-user approach by incorporating gaze and audio features from all participants while keeping the binary VAD specific to the target user. As shown in Fig.~\ref{fig:model architecture} (b), gaze and audio data from each user are processed independently through respective feature extractors, generating individual user feature vectors. These vectors pass through a user attention model, computing attention scores to quantify user interactions~\cite{vaswani2017attention, kim2017structured}. The model outputs feature vectors for each user, concatenated with the target user's binary VAD feature. The final representations go through classification layers. To prevent positional bias, training includes both original and reversed reference user orders.

\section{Experiments and Results}
\subsection{Training setup and evaluation metrics}

The user feature model consists of a 2D CNN block (kernel size: 3, 32 dimensions) followed by a ConvLSTM layer (kernel size: 3, 64 dimensions). The binary VAD model employs a two-layer MLP with 128 and 32 units. We used the Adam optimizer (learning rate: 1e-2) and trained the model using a pre-training and fine-tuning strategy: pre-trained on 6 algorithmically labeled sessions, then fine-tuned on 4 manually labeled sessions. Each manually labeled session was split into first 60\% for training, the middle 20\% for validation, and the final 20\% for testing.

As a baseline, we trained a model using only the target user's binary VAD features, comparing its performance with both the single-user and multi-user models to assess the impact of gaze information on turn-taking prediction. To address class imbalance, we employ a weighted cross-entropy loss function~\cite{aurelio2019learning}. Given the challenges posed by class imbalances, we used the F1 score as the primary metric, applying the macro-averaging method to compute an overall F1 score by averaging per-class scores equally~\cite{opitz2019macro}.

\begin{table}[t]
  \begin{center} 
    \caption{F1 scores for behavior-based classification. \textbf{Bold} marks the best, while \underline{underline} the second-best. (a) Original evaluation method. (a.1) Transition-level analysis: F1 scores computed within the 1-second window following a turn change. (a.2) Group-level analysis: evaluates interaction pattern by analyzing sequences of consecutive data points.}
    \label{table:behavior-based result}
  \resizebox{0.43\textwidth}{!}{
  \begin{tabular}{p{1.5cm} | p{0.8cm} p{0.8cm} p{0.8cm} p{0.8cm} p{0.8cm}}
    \toprule
  \multicolumn{6}{c}{\textbf{(a) Behavior-based}}\\    
    \midrule
    \textbf{Model} & \textbf{All} &\textbf{Turn-taking} & \textbf{Turn-keeping} & \textbf{Back-channel} & \textbf{Silence} \\ 
    \midrule
     Target VAD & 0.661 & 0.490 & 0.730 & 0.424 & \textbf{1.000} \\
     Single-user & \underline{0.670} & \underline{0.518} & \underline{0.735} & \underline{0.429} & \textbf{1.000} \\
     Multi-user & \textbf{0.682} & \textbf{0.534} & \textbf{0.741} & \textbf{0.454} & \textbf{1.000}\\
     \midrule
  \multicolumn{6}{c}{\textbf{(a.1) Transition-level}}\\    
    \midrule
    \textbf{Model} & \textbf{All} &\textbf{Turn-taking} & \textbf{Turn-keeping} & \textbf{Back-channel} & \textbf{Silence} \\ 
    \midrule
     Target VAD & 0.559 & 0.081 & \underline{0.711} & 0.443 & \textbf{1.000} \\
     Single-user & \underline{0.595} & \underline{0.218} & \underline{0.711} & \underline{0.452} & \textbf{1.000} \\
     Multi-user & \textbf{0.653} & \textbf{0.400} & \textbf{0.733} & \textbf{0.481} & \textbf{1.000}\\
    \midrule
  \multicolumn{6}{c}{\textbf{(a.2) Group-level}}\\    
    \midrule
     Target VAD & 0.705 & 0.414 & 0.701 & 0.703 & \textbf{1.000} \\
     Single-user & \underline{0.715} & \underline{0.437} & \underline{0.713} & \underline{0.711} & \textbf{1.000} \\
     Multi-user & \textbf{0.732} & \textbf{0.494} & \textbf{0.720} & \textbf{0.715} & \textbf{1.000} \\
    \bottomrule
  \end{tabular}
  }
  \end{center}
  \vspace{-4mm}
\end{table}

\begin{figure*}[t]
\centering
    \includegraphics[width=0.99\textwidth]{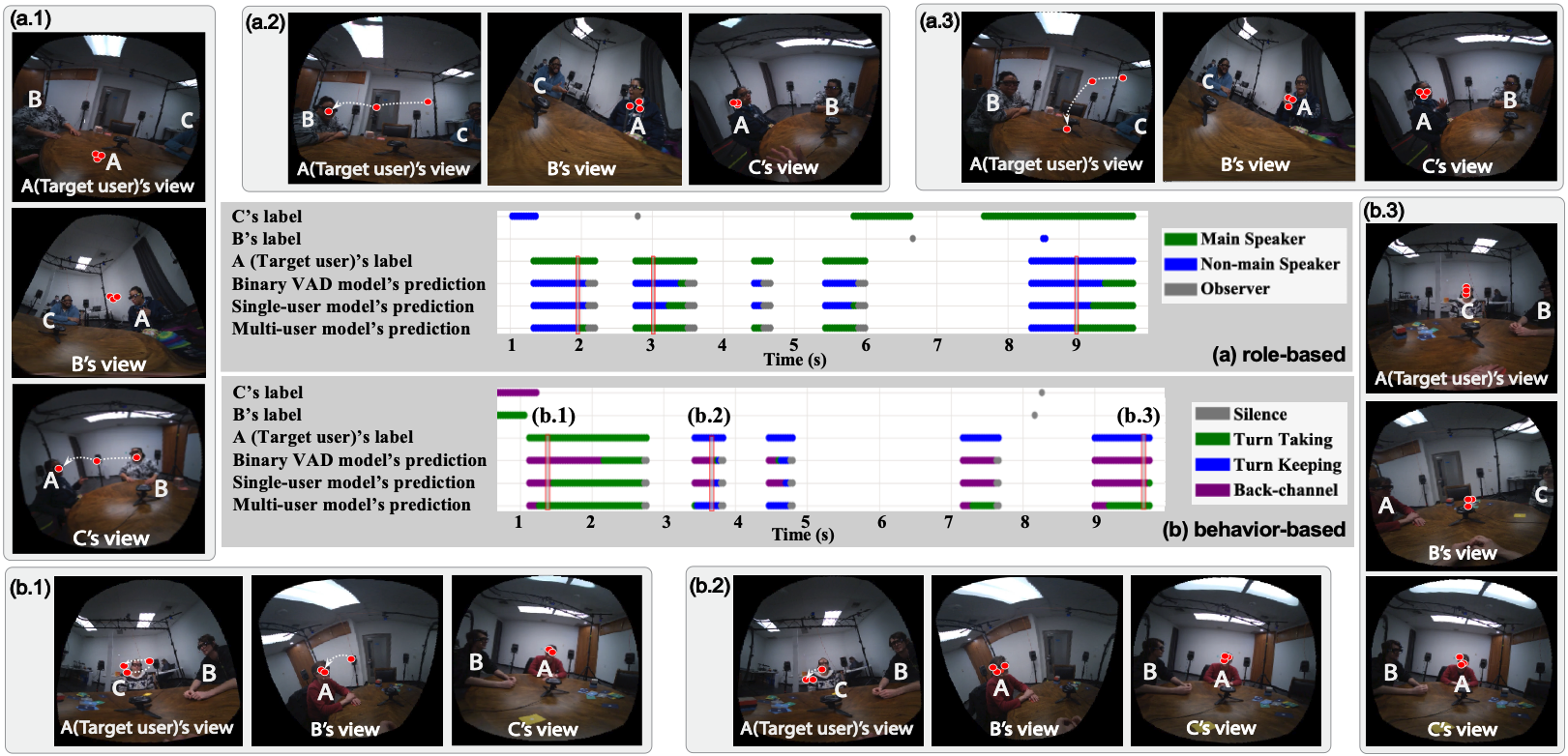}
    \caption{A 10-second video segment illustrating turn-taking behavior for target user A and reference users B and C. Panels (a) and (b) present role-based and behavior-based classifications, respectively, with color-coded turn-taking labels when voice is detected. The first 3 rows display ground truth labels for A, B and C, while the next 3 rows display predictions from the binary VAD, single-user model, and multi-user model for target user A. Gaze direction (red dots) and gaze shifts (white arrows) provide key turn-taking cues. For detailed descriptions of (a.1)–(b.3), please refer to the main text.} 
    \label{fig:video examples}
    \vspace{-3mm}
\end{figure*}

\subsection{Role-based results}

Table~\ref{table:role-based result} shows the F1 scores for role-based classification, comparing the target VAD baseline, single-user model, and multi-user model. Incorporating single-user gaze data improves classification for both main and non-main speakers, while multi-user gaze data further enhances performance. Since the observer role is directly determined by binary VAD indicating silence, the F1 score remains 100\% across all. A gaze-only model without VAD was excluded as it failed to converge, indicating gaze alone is insufficient to distinguish silence periods.

Figure~\ref{fig:confusion matrix} shows confusion matrices. The target VAD-only model often misclassifies main speakers as non-main, while the single-user gaze model reduces errors, and the multi-user model further refines classification for more balanced performance.

\subsection{Behavior-based results}

Table~\ref{table:behavior-based result} summarized the F1 scores for behavior-based classification. Incorporating single-user gaze data improves performance across turn-taking, turn-keeping, and back-channel, with multi-user model further enhancing results.

To provide a detailed evaluation, we analyze transition-level and group-level performance (Table~\ref{table:behavior-based result}). Transition-level analysis assesses predictions within 1s after a turn change, ensuring timely responses~\cite{ishii2016using, wang2024predicting}. Group-level analysis aggregates consecutive data points into single events, capturing broader interaction patterns~\cite{hoppe2016end, elmadjian2023online}. The predicted label for each event is determined by the most frequent class. Both approaches show improvements with single-user gaze data and even greater gains with multi-user gaze data.

\subsection{Analysis of predictions}

To further examine the model’s behavior, we analyze specific cases of turn transitions, comparing the binary VAD, single-user, and multi-user gaze models (Fig.~\ref{fig:video examples}). The following sections highlight key moments from both role-based (panel a) and behavior-based (panel b) classifications, emphasizing the impact of gaze dynamics on predictions.

\subsubsection{Role-based}
\begin{itemize}

\item {\bf (a.1):} Only the multi-user model correctly predicts target user A as the main speaker, likely because both reference users, B and C, direct their gaze toward A. B maintains eye contact with A, while C shifts attention from B to A. However, the correct prediction occurs slightly later, as C's gaze shift is delayed. This joint attention serves as a strong cue, which the binary VAD and single-user models fail to capture.

\item {\bf (a.2):} The single-user model predicts A as the main speaker earlier than the binary VAD model, while the multi-user model identifies it even sooner. A’s gaze aversion, and the B/C's sustained gaze on A enhance the prediction.

\item {\bf (a.3):} The multi-user model appears to underperform, but the issue stems from mislabeling. Around 9 seconds, C starts laughing, unintentionally shifting the conversational turn from C to A. Despite the incorrect label, the multi-user model correctly predicts the transition, capturing nuances that were overlooked in the manual labeling process.
\end{itemize}

\subsubsection{Behavior-based}

\begin{itemize}

\item {\bf (b.1):} The single-user model detects the transition earlier than the VAD model, while the multi-user model identifies it even earlier. A’s gaze aversion, B's shift to A, and C maintaining gaze on A enhances turn-taking detection.

\item {\bf (b.2):} When the ground truth label is turn-keeping, only the multi-user model predicts correctly by leveraging B/C's sustained gaze on A, an indicator of continued engagement.

\item {\bf (b.3):} All models fail to predict turn-keeping, as B disengages by looking away from the speaker, highlighting gaze data’s limitation and the need for additional context.

\end{itemize}

\section{Conclusion}

We introduced a lightweight framework for turn-taking prediction, demonstrating that gaze, despite its inherent complexity, is a reliable predictive cue when structured with spatial constraints. Unlike prior work that overlooked or passively used gaze, our method integrates speaker localization and speech activity to extract meaningful predictive signals without computationally intensive models.

Beyond modeling, this approach is computationally efficient and suited for low-latency applications. It enables adaptive directional sound control based on speaker roles, improving hearing assistance in smart glasses through beamforming.

%This study presented a gaze-enhanced turn-taking prediction model for triadic conversations, utilizing corrected gaze data, spatial heatmaps, and binary VAD features. Our experiments showed that single-user gaze improves prediction accuracy, while the multi-user model further refines classification by capturing richer conversational dynamics.

%By integrating gaze-based cues, our approach provides a lightweight, privacy-conscious solution for applications such as assessing conversation quality, adapting audio levels dynamically based on speaker roles, and enhancing hearing assistance in smart glasses through beamforming. These findings highlight the potential of gaze-driven models for improving multi-party interactions.

%  
%\section{Acknowledgements}
%Acknowledgement should only be included in the camera-ready version, not in the version submitted for review. The 5th page is reserved exclusively for acknowledgements and  references. No other content must appear on the 5th page. Appendices, if any, must be within the first 4 pages. The acknowledgments and references may start on an earlier page, if there is space.

%\ifinterspeechfinal
%     The Interspeech 2025 organisers
%\else
%     The authors
%\fi
%would like to thank ISCA and the organising committees of past Interspeech conferences for their help and for kindly providing the previous version of this template.

\bibliographystyle{IEEEtran}
\bibliography{mybib}

\begin{thebibliography}{10}
\providecommand{\url}[1]{#1}
\csname url@samestyle\endcsname
\providecommand{\newblock}{\relax}
\providecommand{\bibinfo}[2]{#2}
\providecommand{\BIBentrySTDinterwordspacing}{\spaceskip=0pt\relax}
\providecommand{\BIBentryALTinterwordstretchfactor}{4}
\providecommand{\BIBentryALTinterwordspacing}{\spaceskip=\fontdimen2\font plus
\BIBentryALTinterwordstretchfactor\fontdimen3\font minus \fontdimen4\font\relax}
\providecommand{\BIBforeignlanguage}[2]{{%
\expandafter\ifx\csname l@#1\endcsname\relax
\typeout{** WARNING: IEEEtran.bst: No hyphenation pattern has been}%
\typeout{** loaded for the language `#1'. Using the pattern for}%
\typeout{** the default language instead.}%
\else
\language=\csname l@#1\endcsname
\fi
#2}}
\providecommand{\BIBdecl}{\relax}
\BIBdecl

\bibitem{sacks1978simplest}
H.~Sacks, ``A simplest systematics for the organization of turn taking for conversation,'' 1978.

\bibitem{duncan1974structure}
S.~Duncan~Jr, ``On the structure of speaker--auditor interaction during speaking turns1,'' \emph{Language in society}, vol.~3, no.~2, pp. 161--180, 1974.

\bibitem{riest2015anticipation}
C.~Riest, A.~B. Jorschick, and J.~P. de~Ruiter, ``Anticipation in turn-taking: mechanisms and information sources,'' \emph{Frontiers in Psychology}, vol.~6, p.~89, 2015.

\bibitem{degutyte2021role}
Z.~Degutyte and A.~Astell, ``The role of eye gaze in regulating turn taking in conversations: a systematized review of methods and findings,'' \emph{Frontiers in Psychology}, vol.~12, p. 616471, 2021.

\bibitem{jokinen2013gaze}
K.~Jokinen, H.~Furukawa, M.~Nishida, and S.~Yamamoto, ``Gaze and turn-taking behavior in casual conversational interactions,'' \emph{ACM Transactions on Interactive Intelligent Systems (TiiS)}, vol.~3, no.~2, pp. 1--30, 2013.

\bibitem{kendrick2023turn}
K.~H. Kendrick, J.~Holler, and S.~C. Levinson, ``Turn-taking in human face-to-face interaction is multimodal: gaze direction and manual gestures aid the coordination of turn transitions,'' \emph{Philosophical Transactions of the Royal Society B}, vol. 378, no. 1875, p. 20210473, 2023.

\bibitem{hirvenkari2013influence}
L.~Hirvenkari, J.~Ruusuvuori, V.-M. Saarinen, M.~Kivioja, A.~Per{\"a}kyl{\"a}, and R.~Hari, ``Influence of turn-taking in a two-person conversation on the gaze of a viewer,'' \emph{PloS One}, vol.~8, no.~8, p. e71569, 2013.

\bibitem{ekstedt2020turngpt}
E.~Ekstedt and G.~Skantze, ``Turngpt: a transformer-based language model for predicting turn-taking in spoken dialog,'' \emph{arXiv preprint arXiv:2010.10874}, 2020.

\bibitem{chang2022turn}
S.-y. Chang, B.~Li, T.~N. Sainath, C.~Zhang, T.~Strohman, Q.~Liang, and Y.~He, ``Turn-taking prediction for natural conversational speech,'' \emph{arXiv preprint arXiv:2208.13321}, 2022.

\bibitem{kurata2023multimodal}
F.~Kurata, M.~Saeki, S.~Fujie, and Y.~Matsuyama, ``Multimodal turn-taking model using visual cues for end-of-utterance prediction in spoken dialogue systems,'' \emph{Proc. Interspeech 2023}, pp. 2658--2662, 2023.

\bibitem{ishii2016prediction}
R.~Ishii, K.~Otsuka, S.~Kumano, and J.~Yamato, ``Prediction of who will be the next speaker and when using gaze behavior in multiparty meetings,'' \emph{ACM Transactions on Interactive Intelligent Systems (TIIS)}, vol.~6, no.~1, pp. 1--31, 2016.

\bibitem{lee2023multimodal}
M.-C. Lee, M.~Trinh, and Z.~Deng, ``Multimodal turn analysis and prediction for multi-party conversations,'' in \emph{Proceedings of the 25th International Conference on Multimodal Interaction}, 2023, pp. 436--444.

\bibitem{wang2024predicting}
P.~Wang, E.~Han, A.~Queiroz, C.~DeVeaux, and J.~N. Bailenson, ``Predicting and understanding turn-taking behavior in open-ended group activities in virtual reality,'' \emph{arXiv preprint arXiv:2407.02896}, 2024.

\bibitem{jiang2022egocentric}
H.~Jiang, C.~Murdock, and V.~K. Ithapu, ``Egocentric deep multi-channel audio-visual active speaker localization,'' in \emph{Proceedings of the IEEE/CVF Conference on Computer Vision and Pattern Recognition}, 2022, pp. 10\,544--10\,552.

\bibitem{yin2024hearing}
Y.~Yin, I.~Ananthabhotla, V.~K. Ithapu, S.~Petridis, Y.-H. Wu, and C.~Miller, ``Hearing loss detection from facial expressions in one-on-one conversations,'' in \emph{ICASSP 2024-2024 IEEE International Conference on Acoustics, Speech and Signal Processing (ICASSP)}.\hskip 1em plus 0.5em minus 0.4em\relax IEEE, 2024, pp. 5460--5464.

\bibitem{murdock2024self}
C.~Murdock, I.~Ananthabhotla, H.~Lu, and V.~K. Ithapu, ``Self-motion as supervision for egocentric audiovisual localization,'' in \emph{ICASSP 2024-2024 IEEE International Conference on Acoustics, Speech and Signal Processing (ICASSP)}.\hskip 1em plus 0.5em minus 0.4em\relax IEEE, 2024, pp. 7835--7839.

\bibitem{yun2025spherical}
H.~Yun, R.~Gao, I.~Ananthabhotla, A.~Kumar, J.~Donley, C.~Li, G.~Kim, V.~K. Ithapu, and C.~Murdock, ``Spherical world-locking for audio-visual localization in egocentric videos,'' in \emph{European Conference on Computer Vision}.\hskip 1em plus 0.5em minus 0.4em\relax Springer, 2025, pp. 256--274.

\bibitem{engel2023project}
J.~Engel, K.~Somasundaram, M.~Goesele, A.~Sun, A.~Gamino, A.~Turner, A.~Talattof, A.~Yuan, B.~Souti, B.~Meredith \emph{et~al.}, ``Project aria: A new tool for egocentric multi-modal ai research,'' \emph{arXiv preprint arXiv:2308.13561}, 2023.

\bibitem{liu2022pause}
S.~Liu, Y.~Nakajima, L.~Chen, S.~Arndt, M.~Kakizoe, M.~A. Elliott, and G.~B. Remijn, ``How pause duration influences impressions of english speech: Comparison between native and non-native speakers,'' \emph{Frontiers in Psychology}, vol.~13, p. 778018, 2022.

\bibitem{campione2002large}
E.~Campione and J.~V{\'e}ronis, ``A large-scale multilingual study of silent pause duration,'' in \emph{Speech Prosody 2002, International Conference}, 2002.

\bibitem{shi2015convolutional}
X.~Shi, Z.~Chen, H.~Wang, D.-Y. Yeung, W.-K. Wong, and W.-c. Woo, ``Convolutional lstm network: A machine learning approach for precipitation nowcasting,'' \emph{Advances in Neural Information Processing Systems}, vol.~28, 2015.

\bibitem{vaswani2017attention}
A.~Vaswani, ``Attention is all you need,'' \emph{Advances in Neural Information Processing Systems}, 2017.

\bibitem{kim2017structured}
Y.~Kim, C.~Denton, L.~Hoang, and A.~M. Rush, ``Structured attention networks,'' \emph{arXiv preprint arXiv:1702.00887}, 2017.

\bibitem{aurelio2019learning}
Y.~S. Aurelio, G.~M. De~Almeida, C.~L. de~Castro, and A.~P. Braga, ``Learning from imbalanced data sets with weighted cross-entropy function,'' \emph{Neural Processing Letters}, vol.~50, pp. 1937--1949, 2019.

\bibitem{opitz2019macro}
J.~Opitz and S.~Burst, ``Macro f1 and macro f1,'' \emph{arXiv preprint arXiv:1911.03347}, 2019.

\bibitem{ishii2016using}
R.~Ishii, K.~Otsuka, S.~Kumano, and J.~Yamato, ``Using respiration to predict who will speak next and when in multiparty meetings,'' \emph{ACM Transactions on Interactive Intelligent Systems (TiiS)}, vol.~6, no.~2, pp. 1--20, 2016.

\bibitem{hoppe2016end}
S.~Hoppe and A.~Bulling, ``End-to-end eye movement detection using convolutional neural networks,'' \emph{arXiv preprint arXiv:1609.02452}, 2016.

\bibitem{elmadjian2023online}
C.~Elmadjian, C.~Gonzales, R.~L.~d. Costa, and C.~H. Morimoto, ``Online eye-movement classification with temporal convolutional networks,'' \emph{Behavior Research Methods}, vol.~55, no.~7, pp. 3602--3620, 2023.

\end{thebibliography}

\end{document}